\documentclass[twocolumn,showkeys,showpacs,preprintnumbers,prd,superscriptaddress,nofootinbib]{revtex4-1}
\usepackage{graphicx}
\usepackage{epsf}
\usepackage{bm}
\usepackage{amsmath}
\usepackage{amsfonts}
\usepackage{amssymb}
\usepackage{epstopdf}
\usepackage{natbib}
\usepackage{hyperref}
\usepackage{color}
\usepackage{verbatim}
\usepackage{multirow}
\usepackage{bm}
\usepackage{hyperref}
\usepackage{float}
\usepackage{xcolor}
\newcommand{\jcap}{Journal of Cosmology and Astroparticle Physics}
\newcommand{\aap}{Astronomy and Astrophysics}
\newcommand{\apjl}{Astrophysical Journal, Letters}
\newcommand{\mnras}{Monthly Notices of the RAS}

\makeatletter\let\expandableinput\@@input\makeatother

\begin{document}

\title{High-redshift cosmography with a possible cosmic distance duality relation violation}

\author{José F. Jesus}
\email{jf.jesus@unesp.br}
\affiliation{Universidade Estadual Paulista (UNESP), Instituto de Ci\^encias e Engenharia, Departamento de Ci\^encias e Tecnologia - R. Geraldo Alckmin, 519, 18409-010, Itapeva, SP, Brazil}
\affiliation{Universidade Estadual Paulista (UNESP), Faculdade de Engenharia e Ci\^encias, Departamento de F\'isica - Av. Dr. Ariberto Pereira da Cunha 333, 12516-410, Guaratinguet\'a, SP, Brazil}

\author{ Mikael  J. S. Gomes}
\email{mikael.gomes.118@ufrn.edu.br}
\affiliation{Universidade Federal do Rio Grande do Norte, Departamento de F\'isica Te\'orica e Experimental, 59300-000, Natal - RN, Brazil.}

\author{Rodrigo F. L. Holanda}
\email{holandarfl@gmail.com}
\affiliation{Universidade Federal do Rio Grande do Norte, Departamento de F\'isica Te\'orica e Experimental, 59300-000, Natal - RN, Brazil.}
\affiliation{Universidade Federal de Campina Grande, Unidade Acad\^emica de F\'isica, 58429-900, Campina Grande - PB, Brazil.}

\author{Rafael C. Nunes}
\email{rafadcnunes@gmail.com}
\affiliation{Instituto de F\'isica, Universidade Federal do Rio Grande do Sul, 91501-970 Porto Alegre RS, Brazil.}
\affiliation{Divis\~{a}o de Astrof\'{i}sica, Instituto Nacional de Pesquisas Espaciais, Avenida dos Astronautas 1758, S\~{a}o Jos\'{e} dos Campos, 12227-010, S\~{a}o Paulo, Brazil.}

\begin{abstract}

In this study, we used geometric distances at high redshifts (both luminosity and angular) to perform a cosmographic analysis with the Padé method, which stabilizes the behaviour of the cosmographic series in this redshift regime. However, in our analyses, we did not assume the validity of the Cosmic Distance Duality Relation (CDDR), but allowed for potential violations, such as \(d_L(z) = \eta(z)(1+z)^2d_A(z)\), where three different functional forms of \(\eta(z)\) are considered. By incorporating updated data from supernovae (SN), baryon acoustic oscillations (BAO), and cosmic chronometers (CC), we obtained observational constraints on cosmographic models alongside possible CDDR violations. Interestingly, we found that potential CDDR violations introduce new statistical correlations among cosmographic parameters such as \(H_0\), \(q_0\), and \(j_0\). Nonetheless, within this framework, we did not observe significant deviations from the CDDR, and our results remain consistent with the predictions of the \(\Lambda\)CDM model. In the same time, this work provides a novel and straightforward method for testing the CDDR by fixing the background evolution through cosmographic techniques, paving the way for new geometric observational tests of possible deviations from standard cosmology.

\end{abstract}

\maketitle



\section{Introduction} 

Etherington's reciprocity law, established by Etherington in 1933 \cite{1933PMag...15..761E}, is a fundamental principle in cosmology essential for interpreting geometrical distance measurements in astronomical observations. It states that when a source and observer move relative to each other, the solid angles they subtend to one another are linked by geometrical constants that incorporate the redshift $z$ measured by the observer from the source. This law derives from the invariance of various geometric properties when the roles of source and observer are swapped in astronomical observations. Etherington’s reciprocity law can be expressed in various forms, whether in terms of solid angles or through relationships between different cosmological distances. Its most practical form in astronomy, known as the cosmic distance duality relation (CDDR), relates the luminosity distance $D_L$ to the angular diameter distance $D_A$ through the equation \cite{CDDR}

\begin{equation}
\frac{d_L}{d_A (1 + z)^2} = 1.
\label{CDDR1}
\end{equation} 

It is important to stress that the above relation is essential in observational cosmology, and several works have proposed new methods to test its validity. Possible departures from the CDDR could indicate the presence of systematic errors in observations or even the need for new physics \cite{CDDR, BASSET}. In the literature, there are cosmological model-dependent tests \cite{BASSET, 2006IJMPD..15..759D, 2010JCAP...10..024A, 2004PhRvD..70h3533U, 2012JCAP...02..013A, 2012A&A...538A.131H, 2017JCAP...09..039H}, which impose more restrictive limits on possible deviations from (\ref{CDDR1}), while several others are cosmological model-independent and consider only astronomical observations. The basic methodology to test the CDDR has been to consider $\frac{d_L}{d_A (1 + z)^2} = \eta(z)$ and to propose several deformed expressions for $\eta(z)$ \cite{2010ApJ...722L.233H, 2011ApJ...729L..14L, 2012MNRAS.420L..43G, 2012JCAP...06..022H, 2015JCAP...10..061S, 2020PhRvD.102f3513D, 2020EPJP..135..447X, 2020ApJ...892..103Z, 2018MNRAS.480.3117L, 2017JCAP...09..039H, 2016JCAP...04..050R, 2019JCAP...06..008H, 2013PhRvD..87j3530E, 2011RAA....11.1199C, 2021ApJ...909..118Z, 2022EPJC...82..115H, 2020PhRvD.102f3513D}. By considering these works and the progress in the literature, the validity of the CDDR has been verified at least within a $3\sigma$ confidence level. However, given the large errors still present in several observational data, the results have not completely ruled out $\eta \neq 1$ with high statistical significance.

On the other hand, further investigation into cosmic acceleration is imperative to address current challenges. The overlap among different theoretical frameworks for describing dark energy (DE) has spurred the development of methods that do not rely on specific cosmological assumptions \cite{Shafieloo_2006, Sahni_2008, Nesseris_2010, Daly_2003, L_Huillier_2018}. One notable method is cosmography \cite{Visser_2005, Dunsby_2016, D_Agostino_2023, Capozziello_2018_fr, Yang_2020, Lobo_2020}, which uses series expansions of the luminosity distance around the present time, based solely on the cosmological principle. Cosmography offers observable quantities directly comparable to data, independent of assumed DE equations, and has been widely used to distinguish between similar theoretical scenarios \cite{Capozziello_2008, Bamba_2012, Aviles_2012, Capozziello_2017, Capozziello_2022, Rezaei_2020}. Other robust model-independent approaches include Gaussian processes (for a few examples \cite{Seikel_2012,Holsclaw_2010,Sabogal_2024,dinda2024modelagnosticassessmentdarkenergy,jiang2024nonparametriclatetimeexpansionhistory}).

However, cosmography faces challenges: it requires extensive data to distinguish between a cosmological constant ($\Lambda$) and evolving DE, and its reliance on high-redshift data conflicts with its foundation in a Taylor expansion around $z = 0$, leading to convergence issues and error propagation \cite{Busti_2015}. In response, alternative methods have been explored. These include using auxiliary variables and re-parametrizing redshift to ensure convergence at large $z$ \cite{Capozziello_2018, Catto_n_2007, risaliti2018cosmologicalconstraintshubblediagram}. Additionally, rational polynomials like Padé approximations stabilize the cosmographic series' behavior at high redshifts, reducing uncertainties in fitting coefficients and enhancing predictive power \cite{Wei_2014, Gruber_2014, Capozziello_2018_fr, D_Agostino_2023} (See also \cite{Capozziello_2018,Capozziello_2018_fr,Luongo_2020_cub,Capozziello_2019_cub,aviles2017unbiasedestimationsstatefinderparameters} for other relevant works on this topic). Thus, as well established in the literature, standard cosmography often performs poorly when evaluating quantities at high $z$. Addressing this issue requires expanding to higher orders, which introduces numerous free parameters.

In this work, our main aim is to perform a cosmographic expansion on the geometric quantities $d_L(z)$ and $d_A(z)$ in the presence of a CDDR violation, and to constrain the  free parameters associated with possible CDDR violations using geometric measurements from SNe Ia, BAO, and cosmic chronometer samples. We employ the Padé parametrization \cite{Baker_Graves-Morris_1996, Wei_2014, Gruber_2014, Capozziello_2018_fr, D_Agostino_2023}, which allows us to utilize the entire data sample, including points at high redshifts. Actually, the cosmographic parameters will be constrained simultaneously with any possible violation of the CDDR.  Since the data used extends to high redshifts, tight constraints on any violation can be imposed.

The work is organized as follows. In the next section \ref{sec:duality}, we describe the main features of the cosmographic techniques and the $\eta(z)$ parametric models used in our analysis. In Section \ref{sec:data}, we present the methodology and the datasets employed to analyze the model baseline. Section \ref{sec:analysis} presents and discusses our main results, and finally, Section \ref{sec:conclusion} is dedicated to the summary of our findings and conclusive remarks.

\section{\label{sec:duality} Cosmographic Approach and Cosmic Distance  Duality}

\subsection{Cosmographic Approach}The large-scale evolution of the universe can be examined by adhering to the cosmological principle, which posits that the universe is isotropic and homogeneous on the grandest scales. Numerous observations extensively support this principle and leads to the formulation of the Friedmann-Robertson-Walker metric:

\begin{equation}
ds^2 = dt^2 - a(t)^2 \left[dr^2 + r^2 \left( d\theta^2 + \sin^2\theta \, d\varphi^2 \right) \right], \label{eq:FRW}
\end{equation}
where we assume zero spatial curvature. Here, $a(t)$ represents the cosmic scale factor, normalized by $a(t_0) = 1$, with $t_0$ denoting the present time. Throughout this work and the associated discussions, we assume a spatially flat universe by setting $\Omega_k = 0$, as a simplifying hypothesis and as indicated by CMB observations \cite{Planck18}.

Unlike traditional cosmological approaches that rely on solving the Friedmann equations, cosmography provides a kinematic analysis of the universe's expansion, independent of the specific dynamics driving its evolution. Through the cosmographic method, one can derive the history of $a(t)$ directly from observational data, thus circumventing the need to use Einstein's field equations \cite{Visser_2005}.

Several proposals for cosmographic models are available in the literature, particularly focusing on addressing the issues of stability and convergence of geometric distance measures at high redshift $z$ (see \cite{Capozziello_2018,CapozzielloEtAl20} for a review).

Stabilizing cosmographic expansions at high redshift $z$ can also be achieved through the use of rational approximations. These are formulated by taking the ratio of an $n$-th degree polynomial to an $m$-th degree polynomial, creating $(n, m)$-order polynomials that can represent cosmic observables in terms of cosmographic coefficients. The benefit of rational approximations is their flexibility in selecting the appropriate order to enhance the convergence radius, thereby ensuring a stable fitting process.

The advantageous properties of rational polynomials are exemplified by Padé approximations, which have been demonstrated to resolve the high-redshift divergences that affect cosmographic analysis based on Taylor polynomials \cite{Aviles_2014}. The $(n, m)$ Padé approximation of a generic function $f(z)$ is given by \cite{Baker_Graves-Morris_1996}
\begin{equation}
P_{n,m}(z) = \frac{\sum_{i=0}^{n} a_i z^i}{1 + \sum_{j=1}^{m} b_j z^j}, \label{eq:Pade}
\end{equation}
where the sets of coefficients $a_i$ and $b_j$ are determined by matching the $n + m$ derivatives of $P_{n,m}$ evaluated at the origin with the corresponding derivatives from the Taylor series of $f(z)$.

The choice of the appropriate expansion order relates to finding the optimal balance between minimizing the number of free parameters and reducing error propagation in numerical analyses involving data beyond $z \approx 1$. This issue was recently investigated in detail by \cite{CapozzielloEtAl20}, who conducted a comprehensive study using optimization techniques and mathematical considerations to address the degeneracy among coefficients. They concluded that the most suitable Padé approximation for cosmographic applications is the (2,1) order. This approximation yields the following luminosity distance:
\begin{equation}
d_L(z) = \frac{1}{H_0} \left[ \frac{6(q_0-1)z + (-5 - 2j_0 + q_0(8 + 3q_0))z^2}{6(q_0-1) + 2(-1 - j_0 + q_0 + 3q_0^2)z} \right]. \label{eq:luminosity_distance}
\end{equation}
In this case, the Hubble parameter can be obtained as \cite{CapozzielloEtAl20}:
\begin{equation}
H(z) = \frac{2H_0(1+z)^2 (j_0z - 3q_0^2z - q_0(z + 3) + z + 3)}{p_0 + p_1z + p_2z^2},
\label{eq:hubble_parameter}
\end{equation}
where
\begin{align}
p_0 &= 18(-1 + q_0)^2, \\
p_1 &= 6(-1 + q_0)(-5 - 2j_0 + 8q_0 + 3q_0^2), \\
p_2 &= 14 + 7j_0 + 2j_0^2 - 10(4 + j_0)q_0 + (17 - 9j_0)q_0^2 +\nonumber\\
    &+ 18q_0^3 + 9q_0^4.
\end{align}

To ensure the robustness of our results, we will also explore higher-order Padé approximations for the expansion history, specifically the (2,2) and (3,2) cases.

For the (2,2) Padé approximation, the luminosity distance is expressed as:
\begin{widetext}
\begin{align}
\label{dL2}
    d_L(z)&=\frac{1}{H_0}\left(6 z \left(10+9 z-6 q_0^3 z+s_0 z-2 q_0^2 (3+7 z)-q_0 (16+19 z)+j_0 \left(4+\left(9+6 q_0\right)
   z\right)\right)\right)/\nonumber\\
   &\left(60+24 z+6 s_0 z-2 z^2+4 j_0^2 z^2-9 q_0^4 z^2-3 s_0 z^2+6 q_0^3 z (-9+4
   z)+q_0^2 \left(-36-114 z+19 z^2\right)\right.\nonumber\\
   &\left.+j_0 \left(24+6 \left(7+8 q_0\right) z+\left(-7-23 q_0+6
   q_0^2\right) z^2\right)+q_0 \left(-96-36 z+\left(4+3 s_0\right) z^2\right)\right)
\end{align}
\end{widetext}

For the (3,2) order, we have
\begin{widetext}
    \begin{align}
    \label{dL3}
    d_L(z)&=\frac{1}{H_0}(z (-120-180 s_0-156 z-36 l_0 z-426 s_0 z-40 z^2+80 j_0^3 z^2-30 l_0 z^2-135 q_0^6 z^2\nonumber\\
    &-210s_0 z^2+15 s_0^2 z^2-270 q_0^5 z (3+4 z)+9 q_0^4 (-60+50 z+63 z^2)+2 q_0^3
   (720+1767 z\nonumber\\
   &+887 z^2)+3 j_0^2 (80+20 (13+2 q_0) z+(177+40 q_0-60
   q_0^2) z^2)+6 q_0^2 (190+5 (67+9 s_0) z\nonumber\\
   &+(125+3 l_0+58 s_0)
   z^2)-6 q_0 (s_0 (-30+4 z+17 z^2)-2 (20+(31+3 l_0)
   z+(9+4 l_0) z^2))\nonumber\\
   &+6 j_0 (-70+(-127+10 s_0) z+45 q_0^4
   z^2+(-47-2 l_0+13 s_0) z^2+5 q_0^3 z (30+41 z)\nonumber\\
   &-3 q_0^2 (-20+75 z+69 z^2)+2
   q_0 (-115-274 z+(-136+5 s_0) z^2))))/
   (3 (-40-60 s_0\nonumber\\
   &-32z-12 l_0 z-112 s_0 z-4 z^2+40 j_0^3 z^2-4 l_0 z^2-135 q_0^6 z^2-24 s_0 z^2+5 s_0^2 z^2\nonumber\\
   &-30 q_0^5 z
   (12+5 z)+3 q_0^4 (-60+160 z+71 z^2)+j_0^2 (80+20 (11+4 q_0)
   z+(57+20 q_0\nonumber\\
   &-40 q_0^2) z^2)+6 q_0^3 (80+188 z+(44+5 s_0)
   z^2)+2 q_0^2 (190+20 (13+3 s_0) z+(46+6 l_0\nonumber\\
   &+21 s_0) z^2)+4
   q_0 (20+(16+3 l_0) z+(2+l_0) z^2+s_0 (15-17 z-9
   z^2))+2 j_0 (-70\nonumber\\
   &+2 (-46+5 s_0) z+90 q_0^4 z^2+(-16-2 l_0+3
   s_0) z^2+15 q_0^3 z (12+5 z)+q_0^2 (60-370 z\nonumber\\
   &-141 z^2)+2 q_0 (-115-234 z+2
   (-26+5 s_0) z^2))))
\end{align}
\end{widetext}

For the expansions provided in Eqs. \eqref{dL2} and \eqref{dL3}, we compute the expansion rate $H(z)$. Motivated by the arguments above, we will consider Padé approximations to explore the expansion history of the Universe in a model-independent manner. Naturally,  if one considers SNe Ia luminosity distances, $H(z)$ data, and angular diameter distances as provided by BAO measurements, bounds on the free parameters $(H_0, q_0, j_0, l_0, s_0)$ of the equations above can be obtained by assuming the cosmic distance duality relation $d_L(z)=(1+z)^2d_A(z)$.

\subsection{Cosmic distance duality relation}

However, as is widely known, SNe Ia observations are affected by at least four different possible sources of opacity: Milky Way dust, the intergalactic medium, intervening galaxies, or host galaxy opacity. These factors can significantly impact the model parameter estimates and lead to different results. In addition to the possible existence of unexpected new physics, the presence of cosmic opacity directly violates the validity of the CDDR \cite{BASSET, 2010JCAP...10..024A, Avgoustidis_2009, 2013JCAP...04..027H, 2013PhLB..718.1166L, 2015PhRvD..92l3539L, 2018PhRvD..97b3538H, 2022PDU....3701114E, 2020ApJ...899...71L, 2022PDU....3701114E}. Therefore, the validity of the CDDR is not assumed in our analysis. In fact, the cosmographic parameters will be constrained simultaneously with any possible violation of this relation\footnote{The $H(z)$ data used here are obtained from the ages of old passively evolving galaxies and rely only on the detailed shape of the galaxy spectra but not on the galaxy luminosity. At the same time, BAO measurements are completely independent of the measured flux (see Ref.\cite{Avgoustidis_2009} and references therein for more details).}.

By assuming a CDDR violation, equation \eqref{CDDR1} can be written as:
\begin{equation}
d_L(z) = \eta(z) \, d_A(z) \, (1+z)^2,
\end{equation}
while the angular diameter distance $d_A(z)$ can be written as \cite{KumarEtAl22}:
\begin{equation}
d_A(z) = \frac{d_C(z)}{1+z} = \frac{1}{1+z} \int_0^z \frac{dz'}{H(z')}.
\end{equation}

Therefore,
\begin{equation}
d_L(z) = \eta(z) \, (1+z) \int_0^z \frac{dz'}{H(z')}.
\label{dLz}
\end{equation}

It is important to note that in several theoretical contexts where the CDDR can be violated, only the luminosity distance is affected by $\eta(z)$ through Eq. \eqref{dLz} (for a detailed discussion, see, for instance, Ref.~\cite{2014PhRvD..90l4064H}). Therefore, a combination of SNe Ia, CC, and BAO data can yield strong constraints on $\eta(z)$ and the cosmographic parameters, regardless of the CDDR validity assumption. We have chosen to test three CDDR parametrizations \cite{2010ApJ...722L.233H}:
\begin{align}
\label{eta_models}
    \eta_1(z) &= 1 + \eta_0 z, \\
    \eta_2(z) &= 1 + \frac{\eta_0 z}{1+z}. \\
    \eta_3(z) &= (1 +z)^{\eta_0}.
\end{align}

It is important to emphasize that the above expressions are phenomenological and serve to parameterize deviations in the geometric ratio $d_L(z)/[(1+z)^2d_A(z)]$. These parameterizations can be understood as follows:

\begin{itemize}
    \item \textbf{Parameterization 1:} A simple linear correction in cosmic time, designed to quantify deviations at low redshift $z$.
    
    \item \textbf{Parameterization 2:} Inspired by the model \( w(a) = w_0 + (1-a)w_a \) \cite{CHEVALLIER_2001,Linder_2003}, this form provides a smooth parameterization of \( d_L(z)/d_A(z) \), extending to high redshift values.
    
    \item \textbf{Parameterization 3:} Inspired by \cite{Avgoustidis_2009}, this parametric form can phenomenologically account for departures from cosmic transparency. This approach is also based on a Taylor expansion approximation. While the original authors use the notation $\epsilon$, we adopt the equivalent notation $\eta_0$ for consistency with our parameter definitions.

\end{itemize}

Thus, the three parameterizations above provide a degree of theoretical flexibility, allowing us to test our proposal with varying approximations across different $z$ intervals. Our methodology is effectively model-independent, as it relies on basic geometric background quantities—such as the expansion rate and distance measurements—without requiring a fixed $\Lambda$CDM framework, instead utilizing a cosmographic approach.
\\


We highlight that in \cite{2012JCAP...06..022H}, the authors introduced a novel approach based on gas mass fraction measurements in galaxy clusters, employing the $\eta$ model to quantify potential deviations from the CDDR. The authors showed that gas mass fraction measurements obtained via Sunyaev-Zeldovich effect and X-ray observations are linked by $f_{gas}^{SZE}(z)=\eta(z)f_{gas}^{X-ray}(z)$. Notably, since this relation applies to the same objects within a specific galaxy cluster sample, it effectively minimizes potential contamination from systematic errors and redshift differences. Importantly, no specific cosmological framework was adopted in that work.

\section{\label{sec:data} Dataset and Methodology}

In order to determine the constraints on the cosmographic parameter baselines and CDDR free parameters, we consider various combinations of the datasets described below.

\begin{itemize}
\item  The PantheonPlus SNeIa sample \cite{Scolnic_2022}, consisting of 1701 light curves for 1550 distinct SNeIa in the redshift range $0.001 < z < 2.26$. In all but one case, we will consider the uncalibrated PantheonPlus SNeIa sample, whereas in the remaining case we will also consider SH0ES Cepheid host distances, which can be used to calibrate the SNeIa sample \cite{Riess_2022}. We refer to the uncalibrated sample as SN, whereas the SH0ES calibration (when included) is referred to as SN\&SH0ES.

\item Measurements of the expansion rate $H(z)$ from the relative ages of massive, early-time, passively-evolving galaxies, known as cosmic chronometers (CC) \cite{Jimenez_2002}. We use 32 CC measurements in the range $0.07 < z < 1.965$, compiled in Ref. \cite{MorescoEtAl22}. As explained in Ref. \cite{MorescoEtAl22}, this is the first CC compilation with a full estimate of non-diagonal terms in the covariance matrix, and systematics contributions to the latter, arising from effects as: Error in the CC metallicity estimate, error in the CC star formation history, assumption of stellar population synthesis model and rejuvenation effect\footnote{A tutorial to estimate the full CC covariance matrix is provided by the authors of Ref. \cite{MorescoEtAl22} at \url{https://gitlab.com/mmoresco/CCcovariance}}.
We refer to the set of 32 measurements we adopted as CC.

\item  Baryon Acoustic Oscillation (BAO) measurements from surveys such as SDSS and DES have been instrumental in deriving 18 measurements of the angular diameter distance within the redshift range $0.11 < z < 2.4$, as compiled in \cite{Staicova_2022}. All these measurements are calibrated in the form $d_A(z)/r_d$, where $r_d$ represents the sound horizon scale. Given that the $r_d$ scale is essentially insensitive to changes in late-time dynamics, as assumed in this work, we will adopt a specific value $r_d =  147.21 \pm 0.23$ for our main analysis. Different values of $r_d$ parameters have no impact on our results.

\end{itemize}

\begin{table*}
\begin{center}
    \begin{tabular} { l | c | c | c | c }
        \hline
        Parameter & SN+CC & SN\&SH0ES+CC & SN+CC+BAO & SN\&SH0ES+CC+BAO  \\
        \hline
        $M_{\rm B}$     & -                         & $-19.253\pm 0.029$ & -                 & $-19.259\pm 0.028$  \\
$H_0$     & $66.2\pm 4.9$             & $73.2\pm 1.0$      & $65.6\pm 3.7$     & $73.03\pm 0.98$    \\
$q_0$     & $-0.42\pm 0.10$           & $-0.500\pm 0.084$  & $-0.461\pm 0.084$ & $-0.517\pm 0.078$  \\
$j_0$     & $0.71\pm 0.20$            & $0.70\pm 0.19$     & $0.83\pm 0.18$    & $0.87\pm 0.18$ \\
$\eta_0$  & $0.014^{+0.032}_{-0.036}$ & $-0.019\pm 0.024$  & $0.009\pm 0.024$  & $-0.012\pm 0.021$ \\   
        \hline
    \end{tabular}
    \caption{Marginalized constraints on the free parameters of Model 1, expressed as mean values with 68\% confidence intervals, are provided within the framework of the P(2,1) expansion series. These constraints are derived from the various dataset combinations considered in this study.}
    \label{tab1}
\end{center}
\end{table*}

\begin{table*}
\begin{center}
    \begin{tabular} { l | c | c | c | c }
        \hline
        Parameter & SN + CC & SN\&SH0ES + CC & SN + CC + BAO & SN\&SH0ES + CC + BAO \\
        \hline
        $M_{\rm B}$     & -                       & $-19.250\pm 0.029$     & -                       & $-19.256\pm 0.029$ \\
$H_0$     & $65.3\pm 6.0$           & $73.2\pm 1.0$          & $66.6\pm 3.9$           & $73.10\pm 0.98$ \\
$q_0$     & $-0.34^{+0.23}_{-0.30}$ & $-0.60\pm 0.15$        & $-0.52^{+0.14}_{-0.16}$ & $-0.64\pm 0.12$ \\
$j_0$     & $0.62^{+0.30}_{-0.46}$  & $0.97^{+0.33}_{-0.39}$ & $0.96^{+0.29}_{-0.35}$  & $1.16^{+0.29}_{-0.34}$ \\
$\eta_0$  & $0.06^{+0.12}_{-0.15}$  & $-0.081\pm 0.075$      & $0.017\pm 0.073$        & $-0.079\pm 0.059$ \\
        \hline
    \end{tabular}
    \caption{Marginalized constraints on the free parameters of Model 2, expressed as mean values with 68\% confidence intervals, are provided within the framework of the P(2,1) expansion series. These constraints are derived from the various dataset combinations considered in this study.}
    \label{tab2}
\end{center}
\end{table*}

\begin{figure*}[htpb!]
    \centering
    \includegraphics[width=0.49\textwidth]{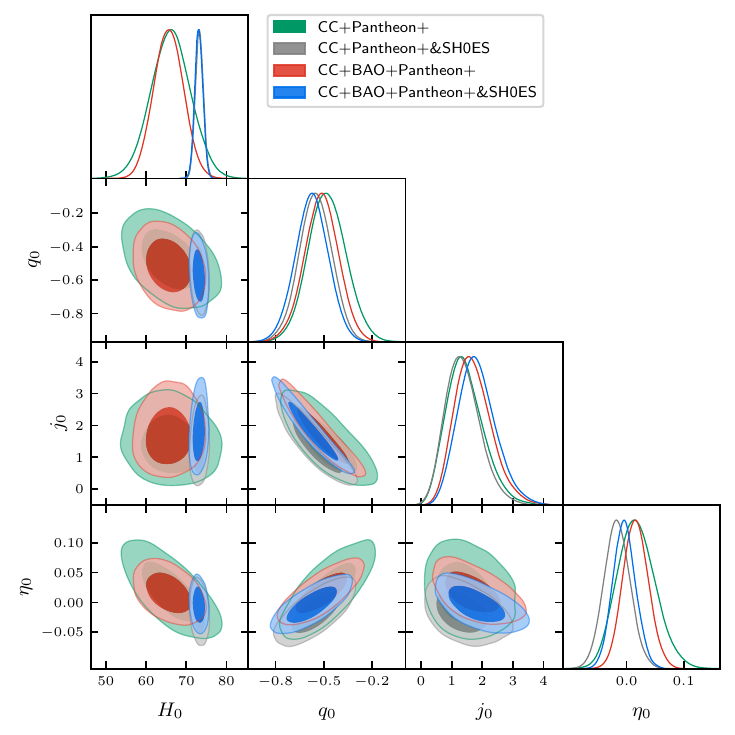} \,\,\,
    \includegraphics[width=0.49\textwidth]{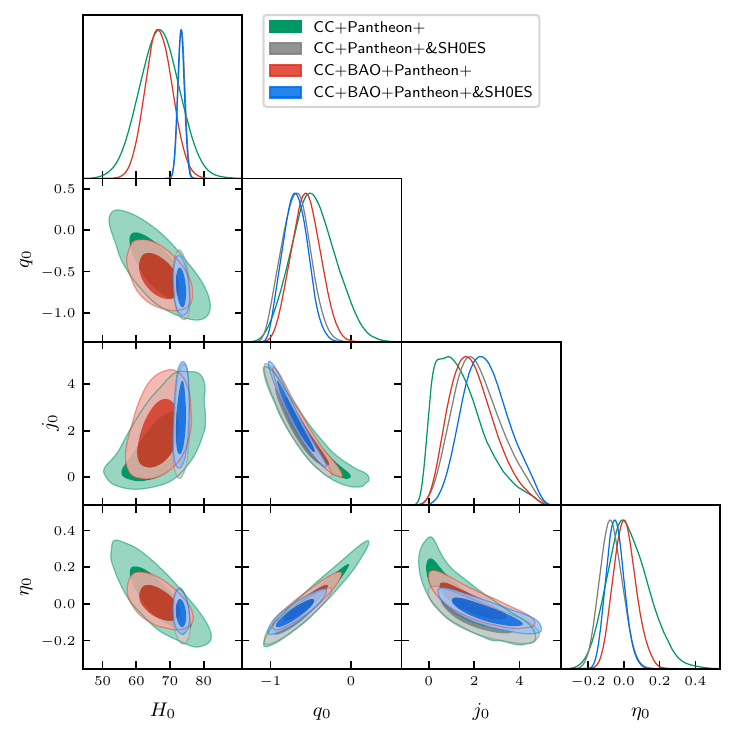}
    \caption{Left panel: 1D posterior distributions and 2D contour plots for Model 1, using different combinations of CC, SN, and BAO samples. Right panel: Same as the left panel, but for Model 2. Both panels from the perspective of the P(2,1) cosmographic expansion.}
    \label{Tplot1}
\end{figure*}

We use the Markov Chain Monte Carlo (MCMC) method to analyze the parameters $\theta_i = \{H_0, q_0, j_0, \eta_0 \}$, building the posterior probability distribution function
\begin{equation}
p(D|\theta) \propto \exp \left( -\frac{1}{2} \chi^2_{\rm total} \right),
\end{equation}
with
\begin{equation}
\chi^2_{\rm total} = \sum_{i=1}^{N} \chi^2_{i},
\end{equation}
where $N$ run the previously described datasets.

The goal of any MCMC approach is to draw $N$ samples $\theta_i$ from the general posterior probability density
\begin{equation}
p(\theta_i, \alpha|D) = \frac{1}{Z} p(\theta, \alpha) p(D|\theta, \alpha),
\end{equation}
where $p(\theta, \alpha)$ and $p(D|\theta, \alpha)$ are the prior distribution and the likelihood function, respectively. Here, the quantities $D$ and $\alpha$ are the set of observations and possible nuisance parameters. The term $Z$ is a normalization factor.

We perform the statistical analysis using the \texttt{emcee} algorithm \cite{Foreman_Mackey_2013}, assuming the theoretical model described in Sec.\ref{sec:duality} and the following priors on the parameter baselines: $H_0 \in [55, 85]$ km/s/Mpc, $q_0 \in [-1, 1]$, $j_0 \in [-5, 5]$, $s_0 \in [-20, 20]$, $l_0 \in [-20, 20]$ and $\eta_0 \in [-1, 1]$. As usual, we discard the first 20\% of the steps of the chain as burn-in.  We follow the Gelman-Rubin convergence criterion, ensuring that all parameters in our chains have $R - 1 < 0.05$, where the parameter $R$ quantifies the Gelman-Rubin statistic, also known as the potential scale reduction factor. In what follows, we discuss our main results. As is standard in state-of-the-art cosmological analyses, our Bayesian inference treats all parameters as free, avoiding any biases associated with fixing their values.

\section{\label{sec:analysis} Results}


We begin our discussion by interpreting the results for the parametric model $\eta_1$, hereafter referred to as model 1. We will use similar nomenclature for the scenario with the parametrization $\eta_2$ and $\eta_3$.

We summarize in Table \ref{tab1} our results at the 68\% CL for the model 1. In all P(2,1) analyses conducted, our baseline model includes four free parameters. Therefore, we consider the combination of SN and CC data as the minimum viable analysis. From a statistical perspective, given the dimensionality of the parameter space, it is not possible to constrain the baseline model using these datasets independently. In Figure \ref{Tplot1} on the left panel, we show the 68\% C.L. and 95\% C.L. contour regions for the cosmographic series obtained from the joint SN + CC, SN\&SH0ES+CC, SN+CC+BAO, and SN\&SH0ES+CC+BAO. It is important to emphasize that adding the SH0ES prior produces an increase in $H_0$, which, in turn, affects the other cosmographic parameters.

For the joint analysis SN + CC, we find $H_0 = 66.2 \pm 4.9$ km/s/Mpc, and we note that $H_0$ is strongly negatively correlated with $\eta_0$. Interestingly, the possible presence of a CDDR violation could create a new statistical correlation in $H_0$. In this sense, we find $\eta_0 = 0.014^{+0.032}_{-0.036}$. Therefore, our constraints on $\eta_0$ are entirely compatible with the null hypothesis, i.e., $\eta_0 = 0$. We note that $j_0 < 1$ at 68\% CL but is entirely compatible with $j_0 = 1$ at any higher statistical significance. Interestingly, $j_0$ is also negatively correlated with $\eta_0$. The presence of $\eta_0$ introduces a new positive correlation with $q_0$. Notably, the presence of the CDDR adds more and stronger statistical correlations with the cosmographic parameters, but the geometric data SN + CC tend to constrain strongly the best-fit values in $\eta_0$, and in return its effects on $H_0$, $j_0$, and $q_0$.

Since SN + CC allows high values in $H_0$, compatible with local measurements within the error bar, we will next consider SN\&SH0ES + CC. For this case, as expected, we note $H_0 = 73.2 \pm 1.0$ km/s/Mpc at 68\% CL. The presence of SH0ES naturally shifts the constraints on $H_0$ to higher values. Due to the strong correlation with $\eta_0$, its value will also naturally shift. In this case, we have $\eta_0 = -0.019 \pm 0.024$ at 68\% CL, inducing a value of $j_0 = 0.70 \pm 0.19$ for the jerk parameter. As expected, the introduction of SH0ES naturally improves the constraints in the model's parameter space. This effect is particularly noted on $\eta_0$.

In our third and fourth rounds of analyses, we introduce BAO data. The main and notable consequence is that in both joint analyses (SN+CC+BAO and SN\&SH0ES+CC+BAO), the jerk parameter is entirely consistent with the $\Lambda$CDM prediction, even at the 68\% CL. Evidently, the other baseline parameters are improved with the inclusion of the BAO sample, as expected. As a result of our more robust joint analysis, specifically using SN\&SH0ES+CC+BAO, we find $\eta_0 = -0.012 \pm 0.02$ at the 68\% CL. This result is fully compatible with the null hypothesis, the $\Lambda$CDM model.

Now, we interpret the results for the model 2. Unlike Model 1, where \(\eta(a)\) exhibits a linear dependence on \(z\), Model 2 introduces a minimal improvement by ensuring that \(\eta(z)\) does not diverge at high \(z\). Since our analysis is limited to low \(z\), we do not anticipate statistically significant differences in the behavior of the function \(\eta(z)\).

In Table \ref{tab2}, we summarize our results at the 68\% CL for Model 2. In Figure \ref{Tplot1} on the right panel, we show the 68\% and 95\% CL contour regions for the cosmographic series obtained from the joint SN + CC, SN\&SH0ES + CC, SN + CC + BAO, and SN\&SH0ES + CC + BAO datasets, as well as for the \(\eta_0\) free parameter.

Due to the similarity in the behavior of the function \(\eta(z)\), all the interpretations discussed above also apply to this model. The only significant difference is that, for this scenario, we find \(\eta_0 = -0.079 \pm 0.059\) at the 68\% CL in our most robust analysis, which is the joint analysis of SN\&SH0ES + CC + BAO. Thus, there is a slight deviation at the 68\% CL indicating a preference for \(\eta_0 < 0\). This trend does not persist at higher significance levels. 

\begin{figure*}[htpb!]
    \centering
    \includegraphics[width=5.9cm]{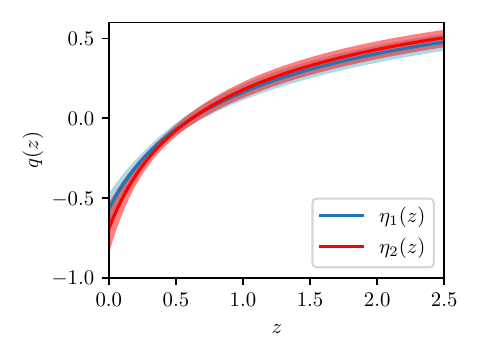}
    \includegraphics[width=5.9cm]{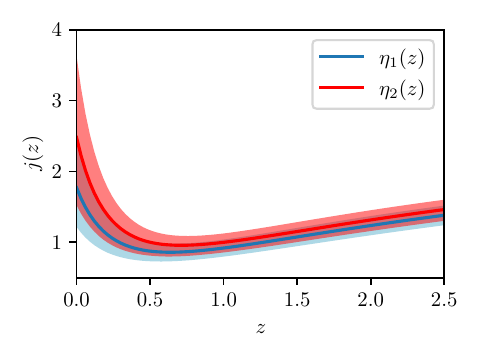}
    \includegraphics[width=5.9cm]{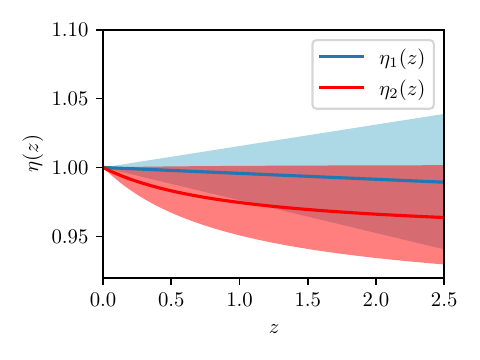}
    \caption{Left Panel: Reconstructions at \(1\sigma\) confidence level of the \(q(z)\) function using the CC + BAO + Pantheon + SH0ES sample. Middle panel: Same as in the left Panel, but for the jerk parameter, $j(z)$. Right Panel: Reconstructions at \(1\sigma\) confidence level of the \(\eta(z)\) function.}
    \label{Eta1Recons}
\end{figure*}

The interpretation of the cosmographic parameters remains consistent with that presented for Model 1. All the conclusions discussed above are the same if we consider Model 3. Therefore, we do not expect that different models of \(\eta(z)\) will have a significant impact on the cosmographic baseline. Conversely, while the cosmographic baseline does exhibit some correlation with \(\eta_0\), the geometric effects are minimal. Thus, we conclude that all results are in good agreement with the predictions of the \(\Lambda\)CDM model.

\begin{table*}[]
\begin{tabular}{|c|c|c|c|c|c|c|c|}
\hline
Model              & $\eta_0$      & $H_0$          & $q_0$                     & $j_0$                  & $s_0$                 & $l_0$               & $\Delta$BIC \\ \hline
$P_{2,1}+\eta_1$   & $-0.004\pm0.020$           & $73.14\pm0.99$ & $-0.57\pm0.10$            & $1.83^{+0.51}_{-0.71}$ & --                    & --                  & 0.67        \\ \hline
$P_{2,1}+\eta_2$   & $-0.048^{+0.047}_{-0.055}$ & $73.3\pm1.0$   & $-0.69^{+0.15}_{-0.17}$   & $2.55^{+0.90}_{-1.2}$  & --                    & --                  & 0.00        \\ \hline
$P_{2,1}+\eta_3$ & $-0.020\pm 0.034$           & $73.2\pm 1.0$  & $-0.62\pm 0.13$           & $2.13^{+0.65}_{-0.94}$ & --                    & --                  & 0.48        \\ \hline
$P_{2,2}+\eta_1$   & $-0.015\pm0.022$           & $73.03\pm0.99$ & $-0.53\pm0.11$            & $1.03^{+0.59}_{-1.1}$  & $1.98^{+0.63}_{-4.4}$ & --                  & 5.35        \\ \hline
$P_{2,2}+\eta_2$   & $-0.086\pm 0.058$          & $73.2\pm 1.0$  & $-0.69\pm 0.16$           & $1.82^{+0.86}_{-1.4}$  & $5.9^{+1.8}_{-8.2}$   & --                  & 4.20        \\ \hline
$P_{2,2}+\eta_3$   & $-0.044\pm 0.040$ & $73.1\pm 1.0$  & $-0.59\pm 0.13$ & $1.28^{+0.68}_{-1.1}$  & $2.9^{+1.0}_{-5.3}$   & --                  & $4.77$ \\ \hline
$P_{3,2}+\eta_1$   & $-0.013\pm 0.021$          & $72.9\pm 1.0$  & $-0.444^{+0.17}_{-0.093}$ & $0.2^{+2.3}_{-1.9}$    & $-7.5^{+3.7}_{-12}$   & $-3.2^{+9.4}_{-26}$ & 10.58       \\ \hline
$P_{3,2}+\eta_2$   & $-0.063\pm 0.056$          & $73.0\pm 1.0$  & $-0.57\pm 0.19$           & $0.5\pm 1.8$           & $-4^{+12}_{-16}$      & $-4.1^{+9.1}_{-25}$ & 10.02       \\ \hline
$P_{3,2}+\eta_3$   & $-0.034\pm 0.038$ & $73.0\pm 1.0$  & $-0.50^{+0.18}_{-0.13}$ & $0.0^{+2.3}_{-2.0}$ & $-7^{+10}_{-13}$ & $-3.6^{+9.1}_{-25}$ & 11.31       \\ \hline
\end{tabular}
\caption{Constraints at 68\% CL are presented for each parameterization under consideration in this work, based on the joint analysis of SN\&SH0ES + CC + BAO data. For comparison, we also report the \(\Delta\mathrm{BIC}\) values for each analysis.}
\label{bics}
\end{table*}

To validate and further solidify the results discussed above, and to examine potential deviations from our assumptions, we will now consider higher-order analyses in our cosmographic expansion, using equations (\ref{dL2}) and (\ref{dL3}) as well as extending our analyses using the parameterization \(\eta_3\) (Model 3). For this purpose, we will only consider the joint analysis of SN\&SH0ES + CC + BAO, as our parameter space will be now larger. These datasets are consistent with each other within our cosmographic framework. Then, without loss of generalization, we will directly consider SN\&SH0ES + CC + BAO.

\begin{figure*}[htpb!]
    \centering
    \includegraphics[width=0.65\textwidth]{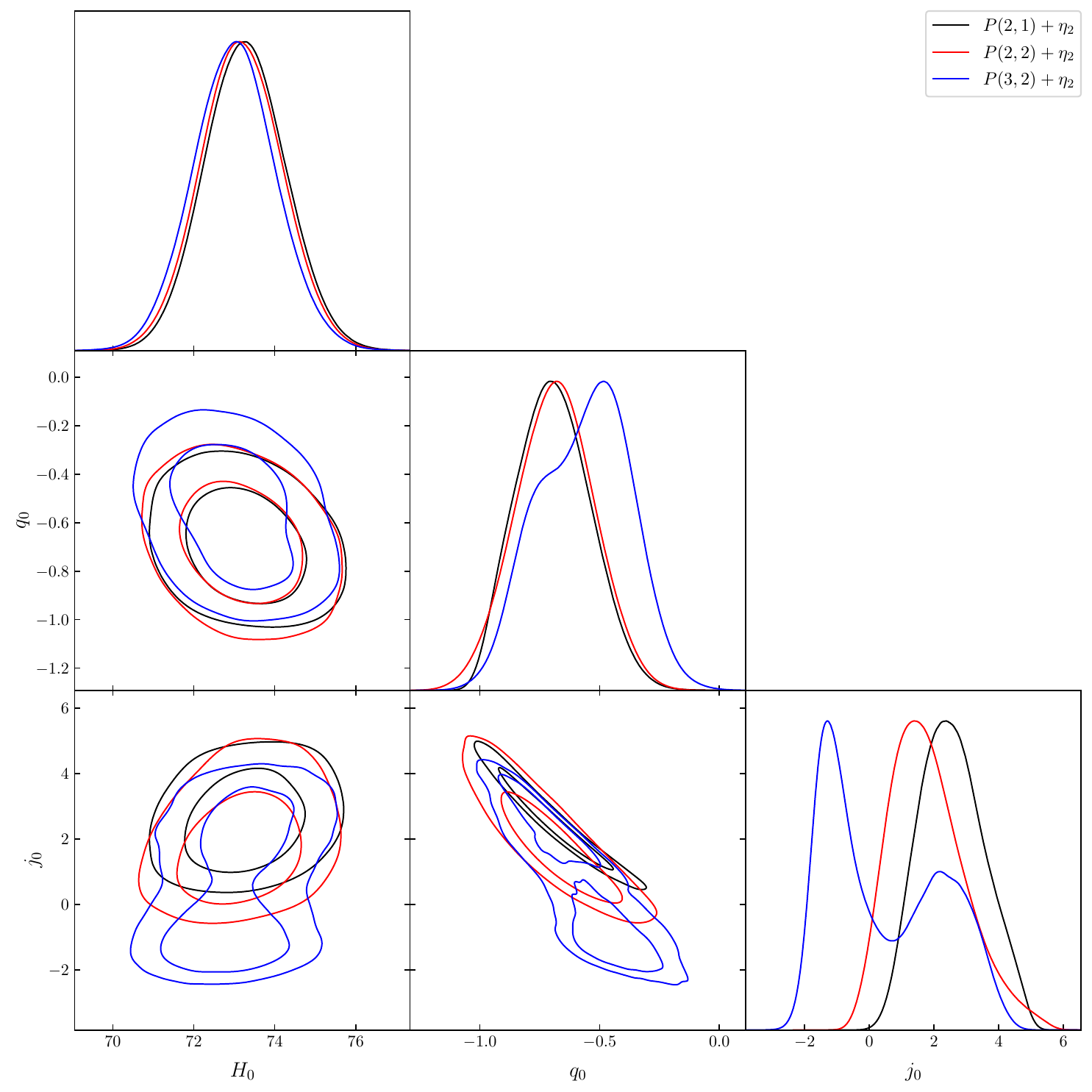}
    \caption{1D posterior distributions and 2D contour plots for all cosmographic expansions considered in this work, based on the joint analysis of SN+SH0ES, CC, and BAO data. Results correspond to the CDDR parametrization for Model 2 (the $\eta_2$ parametrization).}
    \label{Tplot3}
\end{figure*}

Table \ref{bics} summarizes our main results at the 68\% CL for all analyzed cases. The notations P(2,1), P(2,2), and P(3,2) represent different orders of the cosmographic expansion, while $\eta_1$, $\eta_2$, and $\eta_3$ correspond to the parameterizations defined for Models 1, 2, and 3, respectively. Figure \ref{Tplot3} presents the one-dimensional posterior distributions and two-dimensional contour plots for Model 2, which showed the lowest BIC\footnote{See explanation below.} value among the models. A similar conclusion and corresponding plots are observed for the other models.

The constraints on the first parameters of the cosmographic expansion, namely $H_0$ and $q_0$, are consistent across all analyses performed, with no significant differences observed. For the next parameter, $j_0$, the best-fit values exhibit some variation between models; however, the large error bars result in all constraints being compatible at the 95\% CL. For instance, in the case of the three $\eta(z)$ models analyzed, we find $j_0 > 1$ at 68\% CL when interpreted within the cosmographic model P(2,1). This significance, however, diminishes in the contexts of P(2,2) and P(3,2). Consequently, these apparent deviations at 68\% CL can be regarded as potential statistical fluctuations, as the trend is not consistently observed across other cosmographic scenarios.

The parameters associated with higher-order corrections in the cosmographic expansion, such as $s_0$ and $l_0$, are primarily constrained by the priors. This result is expected, given that our data sample lacks the statistical strength required to tightly constrain models with numerous free parameters. Consequently, the constraints on the cosmographic parameters $H_0$, $q_0$, $s_0$, and $l_0$ remain consistent across all models analyzed even at 68\% CL, and $j_0$ at 95\% CL.

For $\eta_0$, we find that its constraints are independent of both the expansion order and the functional form of $\eta(z)$. Thus, the P(2,1) model for the cosmographic expansion demonstrates robustness for our purposes. As the most economical model, with fewer free parameters, we recommend adopting this approach for future cosmographic analyses.

To better quantify the agreement (or disagreement) between the models, we perform a statistical comparison of the models with the minimal \(P_{2,1}\) scenario using the well-known Bayesian Information Criterion (BIC) \cite{Schwarz78,Liddle04}.  The BIC is defined as:
\begin{equation}
\text{BIC} =  k \ln(N) - 2 \ln(\mathcal{L})
\end{equation}
where  $\mathcal{L}$ is the maximized value of the likelihood function of the model, $N$ is the number of data points in the joint analysis and $k$ is the number of parameters estimated by the model.

From these definitions, we calculated
\begin{equation}
\Delta \mathrm{BIC}_i = \mathrm{BIC}_i - \mathrm{BIC}_{\mathrm{model}},
\end{equation} 
where the (2,1) expansion scenario for the $\eta_2$ parametrization serves as our base model. The index $i$ runs over all other possibilities. Given that $\Delta$BIC is an approximation to the Bayes factor $B_{ij}$ as $\Delta\text{BIC}_{ij}=-2\ln B_{ij}$ \cite{SzydlowskiEtAl15}, we may adopt the following criteria, based on Jeffreys scale \cite{Jeffreys39}:
\begin{itemize}
    \item $\Delta \mathrm{BIC} \leq 2$: Barely worth mentioning  
    \item $2 < \Delta \mathrm{BIC} \leq 5$: Substantial evidence  
    \item $5 < \Delta \mathrm{BIC} \leq 7$: Strong evidence
    \item $7 < \Delta \mathrm{BIC} \leq 9$: Very strong evidence  
    \item $\Delta \mathrm{BIC} > 9$: Decisive evidence  
\end{itemize}

Therefore, when comparing with the values shown in Table \ref{bics}, it becomes evident that higher-order cosmographic polynomials provide stronger evidence in favour of the lower-order expansions. In particular, these criteria means that there is substantial to strong evidence against fourth-order (P(2,2)) expansions and decisive evidence against fifth-order (P(3,2)) expansions. This result quantifies that a more extended expansion in the Padé series leads to a better fit to the data relative to lower-order expansions. However, this does not imply significant changes in the observational constraints across the full set of model parameters, as previously discussed.

Figure \ref{Eta1Recons} presents the statistical reconstructions, including \(1\sigma\) and \(2\sigma\) confidence intervals, for the functions \(q(z)\), \(j(z)\), and \(\eta(z)\) in the $\eta_1$ and $\eta_2$ scenarios investigated in this paper. As discussed, no significant differences were found between the studied models. The reconstructions were obtained using standard error propagation techniques, accounting for the appropriate correlations between all baseline parameters. The $\eta_3$ model exhibits similar behaviour as both models at a statistical level.

On the other hand, it is also important to interpret our results in the context of other recent studies in the literature. Cosmographic constraints from the recent DESI-BAO samples \cite{desicollaboration2024desi2024vicosmological} are investigated in \cite{luongo2024modelindependentcosmographicconstraints}, which report a significant departure in $j_0$ at the 1$\sigma$ CL, although the results remain compatible with the $\Lambda$CDM paradigm at higher statistical significance. Similarly, \cite{pourojaghi2024lambdacdmmodelcosmographypossible}, also using DESI-BAO data, shows that $j_0$ may significantly deviate from $\Lambda$CDM predictions. It is worth noting that our analyses do not incorporate the latest DESI-BAO data, which appear to suggest deviations in $j_0$ when this dataset is used. Nevertheless, our results are consistent with $j_0 = 1$ across all analyses performed, irrespective of the cosmographic order.

\section{\label{sec:conclusion} Conclusion}

In this paper, we conducted a comprehensive analysis of the CDDR and its potential violations using cosmography with the Padé parametrization at various orders of expansion. This parametrization is particularly advantageous as it allows the incorporation of observational data even at high redshifts. We explored three possible models for CDDR violation, denoted as $\eta_1$, $\eta_2$ and $\eta_3$. In all our analyses, the cosmographic parameters—$H_0$, $q_0$, and $j_0$—were well simultaneously constrained.

The analysis across all parametric models shows strong consistency with the null hypothesis, providing no compelling evidence for violations of the CDDR. For the $\eta_1$ model, the cosmographic parameters, particularly $H_0$ and $\eta_0$, exhibited significant correlations, with results consistently aligning with the $\Lambda$CDM model, even after incorporating additional data such as SH0ES and BAO (see Fig. 1). In the case of the $\eta_2$ model, a slight preference for $\eta_0 < 0$ was observed at the 68\% CL, although this trend does not persist at higher confidence levels (see Fig. 2). Extensive analyses were conducted using different cosmographic orders, revealing no statistically significant differences in the baseline model parameters. A summary of all results is presented in Table \ref{bics}.

The analysis of the cosmographic parameters $q_0$ and $j_0$ revealed important correlations: $q_0$ showed a positive correlation with $\eta_0$, suggesting that a potential CDDR violation could impact the interpretation of the universe's acceleration, while $j_0$ remained consistent with the predictions of the $\Lambda$CDM model, particularly with the inclusion of BAO data, further reinforcing the robustness of the standard model. In both models, the inclusion of BAO data further constrained the parameters without indicating substantial deviations from the standard cosmological model.

Thus, our analyses support the null hypothesis and the predictions of the $\Lambda$CDM model, with no significant evidence of CDDR violations. As a future perspective, it would be valuable to consider data at higher redshifts to assess the impact on the cosmographic parameters, as well as on the parametric functions $\eta(z)$.
\\

\begin{acknowledgments}
The authors express their gratitude to the referee for their valuable comments and suggestions, which have greatly enhanced the overall quality of the work. The authors also express their gratitude to Rocco D'Agostino for his valuable discussions and suggestions on cosmographic models. J.F.J. thanks the financial support by CNPq under the project No. 314028/2023-4. R.C.N. thanks the financial support from the CNPq under the project No. 304306/2022-3, and the Fundação de Amparo à Pesquisa do Estado do RS (FAPERGS, Research Support Foundation of the State of RS) for partial financial support under the project No. 23/2551-0000848-3. RFLH thanks the financial support from the CNPq under the project No. 308550/2023-47.
\end{acknowledgments}

\bibliography{Refs}

\end{document}